\documentclass[aps, prd, reprint, superscriptaddress, amsmath, amssymb, nofootinbib, floatfix]{revtex4-2}
\usepackage{graphicx}
\usepackage{bm}
\usepackage{multirow}
\usepackage{url}
\usepackage{natbib}
\usepackage{physics}
\usepackage[table,xcdraw]{xcolor}
\usepackage[caption=false]{subfig}
\usepackage{ulem}
\usepackage{booktabs}
\usepackage[colorlinks=true, citecolor=blue, urlcolor=blue, linkcolor=blue]{hyperref}
\definecolor{violet}{rgb}{102, 0, 153}

\begin{document}
\title{Observable Signatures of a Quarkyonic Phase in Neutron Stars}
\author{Probit J Kalita}
\affiliation{Department of Physics \& Astronomy, National Institute of Technology, Rourkela 769008, India}

\author{Tuhin Malik}
\affiliation{CFisUC, Department of Physics, University of Coimbra, PT 3004-516 Coimbra, Portugal}

\author{Tianqi Zhao}
\affiliation{Institute for Nuclear Theory, University of Washington, Seattle, WA~98195, USA}
\affiliation{Network for Neutrinos, Nuclear Astrophysics, and Symmetries (N3AS), University of California, Berkeley, Berkeley, CA~94720, USA}

\author{Bharat Kumar}
\email{kumarbh@nitrkl.ac.in}
\affiliation{Department of Physics \& Astronomy, National Institute of Technology, Rourkela 769008, India}

\author{James M. Lattimer}
\affiliation{Department of Physics \& Astronomy, Stony Brook University, Stony Brook, NY 11794, U.S.A.}

\date{\today}

\begin{abstract}
    Quarkyonic matter in \(\beta\)-equilibrium is a potential description of cold dense matter in neutron stars (NSs), that introduces non-interacting quarks alongside nucleons and leptons in NS cores.
    In this paper, we impose observational and theoretical constraints on the model to perform Bayesian inference on it, and find that it is possible to have quarkyonic matter equations of state that satisfy all current astrophysical observations, thereby reinforcing the argument for its use alongside traditional ones.
    To differentiate between NSs where a quarkyonic phase does and does not appear in the core, we identify some novel signatures based on the mass-radius relation.
    Focusing on canonical (\(1.4\ \mathrm{M_\odot}\)) NSs, we find the populations of NSs with and without quarkyonic cores show separability on the basis of the slope and curvatures of the mass-radius curve, the central sound speed of the star, and the radius difference between two NSs of \(2\ \mathrm{M_\odot}\) and \(1.4\ \mathrm{M_\odot}\).
    Our results indicate that observing a neutron star with these signatures matching the values for quarkyonic core NSs would provide a strong evidence for the existence of a quarkyonic phase or a similar crossover transition in its core.
\end{abstract}

\maketitle

\section{Introduction}
The composition of neutron stars (NSs)- relics of what were once massive stars- and a universal description of matter in the extreme densities inside them remain elusive mysteries.
The interesting situation which we find ourselves in is accommodating two constraints from astronomical observations, the $\gtrsim 2\ \mathrm{M_\odot}$ observations of neutron star mass~\cite{Demorest:2010bx, Antoniadis:2013pzd}, and the compact radii inferred from the binary neutron star merger event GW170817~\cite{LIGOScientific:2018cki}.
These indicate that matter in the NS core would be at very high pressures in the deep interior to allow for the high mass, while also indicating that the pressure towards the core's exterior would be relatively low to allow for the high mass to get compressed into a $\lesssim 13.5\ \mathrm{km}$ radius~\cite{drischler2021limiting}.
This corresponds to the speed of sound ($c_s = \sqrt{\pdv*{P}{\mathcal{E}}}$, where $P$ is pressure and $\mathcal{E}$ is energy density of the matter), the rate at which a perturbation would travel mechanically in the star, increasing rapidly with increasing density in the core~\cite{Tews:2018kmu, McLerran:2018hbz}.
Recent multimessenger inferences of the neutron-star sound speed~\cite{Brandes:2022nxa, Biswas:2024hja} have reinforced the need for microscopic signatures capable of linking $c_s^2(r)$ to directly observable quantities.

In quarkyonic matter, first identified in the large-$N_c$ limit~\cite{McLerran:2007qj}, nucleons emerge as color-singlet quasiparticles confined to a shell near the Fermi surface~\cite{McLerran:2018hbz}.
Deep inside the Fermi sea, quarks fill momentum states and dominate the bulk thermodynamics while remaining globally confined~\cite{Hidaka:2008yy}.
This distinct structure of the Fermi sea arises from the duality between hadronic and quark degrees of freedom~\cite{Fujimoto:2023mzy}, naturally resolving the hyperon puzzle~\cite{Fujimoto:2024doc}.
Improvements to the model were made by Ref.~\cite{Zhao:2020dvu}, where protons and leptons were incorporated into the system in a manner that ensured the NS matter remains in chemical equilibrium.
This description of NS matter allowed for observational constraints to be satisfied while also ensuring that $c_s^2$, after its rapid rise at NS core densities, would decrease to the conformal limit of $1/3$ at high densities.

Given this new theoretical framework for generating the NS matter equation of state (EoS), two pertinent questions arise: \textit{How well constrained are the model parameters, and how well can they be constrained? And, can one possibly distinguish NSs with a quarkyonic phase from a completely hadronic NS?}\@
To address these questions we perform a Bayesian inference of the model parameters using multimessenger observations and theoretical limits as constraints, improving the understanding and accuracy of the model.
Using the slope of the mass-radius relation ($\dv*{R}{M}$), the curvature of the $R(M)$ relation ($\kappa_R$), and the difference in radii of $2.0\ \mathrm{M_\odot}$ and $1.4\ \mathrm{M_\odot}$ stars -which have been used to point towards potential signatures of hyperons in NSs~\cite{Ferreira:2025dat, Bauswein:2025dfg}- our results show that these quantities and $c_s^2$ have distinguishable change in distribution when the quarkyonic phase appears, which can be used in identifying the presence of a quarkyonic phase in the star.
While reductions in the slope and curvature of the mass-radius relation have previously been associated with EOS softening due to hyperons, similar behaviour may also arise from other softening mechanisms, including a first-order phase transition to quark matter.
In contrast, we find that an enhanced slope and curvature in the mass-radius relation results from the rapid EOS-stiffening characteristic of quarkyonic matter.
We will refer to this behaviour as a quarkyonic-like signature because it arises naturally from the emergence of quarkyonic matter, although other mechanisms capable of generating similarly abrupt stiffening may lead to comparable observational features.
Thus, the signature is indicative of, but not uniquely attributable to, a quarkyonic phase.

We employ geometrized units $G=c=1$ throughout this paper.

\section{Methods}
\subsection{The \(\beta\)-equilibriated Quarkyonic Matter Model}
We employ the modified quarkyonic model presented by Ref.~\cite{Zhao:2020dvu}, which includes protons and leptons in \(\beta\)-equilibrium.
This model assumes that strongly interacting quarks near the Fermi surface form interacting nucleons in triplets, while the remaining quarks in the Fermi sea are non-interacting.
The leptons in the model are also taken to be non-interacting fermions.

The interactions between nucleons is described using a nuclear potential energy that is dependent on the number densities of both neutrons (\(\rho_n\)) and protons (\(\rho_p\)), with the potential taking the form~\cite{Zhao:2020dvu}
\begin{align}
	V (\rho_n, \rho_p) = {} & 4x (1-x) (a_0 u + b_0 u^\gamma) \nonumber     \\
	                        & + \pqty{1 - 2x}^2 (a_1 u + b_1 u^{\gamma_1}).
	\label{eq:1_nuclear_interaction_potential}
\end{align}
Here \(u = (\rho_n + \rho_p) / \rho_0\) and \(x = \rho_p / (\rho_n + \rho_p)\), with \(\rho_0 = 0.16\ \mathrm{fm^{-3}}\) being the nuclear saturation density.
The remaining terms, \(a_0,\ a_1,\ b_0,\ b_1\) and \(\gamma\) are determined via fittings to symmetric and pure nuclear matter (SNM, PNM), and are expressed as
\begin{align}
	\gamma = \frac{K/9 - T''_{1/2}}{T_{1/2} - T'_{1/2} + B}, \quad & b_0 = \frac{K/9 - T''_{1/2}}{\gamma(\gamma - 1)}, \nonumber    \\
	a_0 = - B - T_{1/2} - b_0, \quad                               & b_1 = \frac{L/3 + B - J + T_0 - T'_0}{\gamma_1 - 1}, \nonumber \\
	a_1 = J - B - T_0 - b_1. \quad                                 & {}
	\label{eq:2_interaction_parameters}
\end{align}
\(T_{1/2} \simeq 21.79\ \mathrm{MeV}\), is the SNM kinetic energy, while \(T_{1/2}' = u \dv*{T_{1/2}}{u} \simeq 14.34\ \mathrm{MeV}\) and \(T_{1/2}'' = u^2 \dv*[2]{T_{1/2}}{u} \simeq -5.030\ \mathrm{MeV}\) are its first two logarithmic derivatives evaluated at \(\rho_0\).
\(T_0 \simeq 34.33\ \mathrm{MeV}\) and \(T_0' = u \dv*{T_0}{u} \simeq 22.41\ \mathrm{MeV}\) are the PNM kinetic energy and its first logarithmic derivative evaluated at \(\rho_0\).
This leaves us with the free parameters in the nuclear interaction potential, which are the symmetry energy (\(J\)), slope of symmetry energy (\(L\)), incompressibility (\(K\)), binding energy (\(B \equiv BE/A\)) and \(\gamma_1\).

At low densities, up to a transition density \((\rho_t)\) between hadronic and quarkyonic matter, all quarks remain confined within nucleons and the entire Fermi sea is occupied by these nucleons.
Beyond \(\rho_t\), we have quarkyonic matter where nucleons are restricted to the surface of the Fermi sea above a minimum allowed momentum \(k_{0(n,p)}\) given by
\begin{equation}
	k_{0(n,p)} = \pqty{k_{F(n,p)} - k_{t(n,p)}} \bqty{1 + \frac{\Lambda_{QCD}^2}{k_{F(n,p)} k_{t(n,p)}}}.
	\label{eq:3_minimum_nucleon_momenta}
\end{equation}
\(\Lambda_{QCD}\) is the QCD scale parameter, which is also considered a free parameter of the model.
The transition Fermi momenta, \(k_{t(n,p)}\), are the values of the Fermi momenta \(k_{F(n,p)}\) from the \(\beta\)-equilibriated \(n,\ p,\ e,\ \mu\) matter at \(\rho_t\).
The nucleon number densities in the quarkyonic matter are thus
\begin{equation}
	\rho_{(n,p)} = \frac{(k_{F(n,p)}^3 - k_{0(n,p)}^3)}{3\pi^2}.
	\label{eq:4_nucleon_number_density}
\end{equation}

The energy density of the interacting nucleons is then given by
\begin{align}
	\mathcal{E}_B = {} & \sum \limits_{i=n,p} \frac{1}{\pi^2} \int \limits_{k_{0i}}^{k_{Fi}} k^2 \sqrt{m_i^2 + k^2} \dd{k} \nonumber \\
	                   & + (\rho_n + \rho_p) V(\rho_n, \rho_p).
	\label{eq:5_nucleon_energy_density}
\end{align}

The baryon number and charge conservation equations in quarkyonic matter take the forms
\begin{align}
	0 = {} & \rho_B - \qty(\rho_n + \rho_p + \frac{\rho_u + \rho_d}{3}),    \\
	0 = {} & \rho_p + \frac{2\rho_u - \rho_d}{3} - \qty(\rho_e + \rho_\mu),
	\label{eq:6_7_baryon_and_charge_conservation}
\end{align}
with the lepton number densities given by
\begin{equation}
	\rho_{e, \mu} = \frac{k_{F(e,\mu)}^3}{3\pi^2},
	\label{eq:8_lepton_number_density}
\end{equation}
and the quark number densities are given by
\begin{equation}
	\rho_{u, d} = \frac{k_{F(u,d)}^3}{\pi^2}.
	\label{eq:9_quark_number_density}
\end{equation}
The chemical potentials of the quarks are obtained by imposing the condition of chemical equilibrium between nucleons and quarks, to get
\begin{equation}
	\mu_d = \frac{2}{3}\mu_n - \frac{1}{3}\mu_p, \quad \text{and} \quad \mu_u = \frac{2}{3}\mu_p - \frac{1}{3}\mu_n.
	\label{eq:10_quark_chemical_potential}
\end{equation}
Here \(\mu_{n,p}\) are the nucleon chemical potentials,
\begin{equation}
	\mu_{n,p} = \dv{\mathcal{E}_B}{\rho_{n,p}}.
	\label{eq:11_nucleon_chemical_potential}
\end{equation}
The evaluation of these derivatives requires the implicit dependence of
$k_{0(n,p)}$ and $k_{F(n,p)}$ on $\rho_{n,p}$, where
$k_{0(n,p)}$ and $k_{F(n,p)}$ are defined by
Eqs.~(\ref{eq:3_minimum_nucleon_momenta}) and
(\ref{eq:4_nucleon_number_density}), respectively.

The \(\beta\)-equilibrium condition for charge neutral matter can be finally imposed by the following relation between the chemical potentials of the components of the system,
\begin{equation}
	\mu_n - \mu_p = \mu_e = \mu_\mu = \mu_d - \mu_u.
	\label{eq:12_beta_equilibrium}
\end{equation}

Since both \(u\)- and \(d\)-quarks appear at the same transition density, the quark masses are not free model parameters, but are instead obtained from requiring that quarks have zero Fermi momenta at \(\rho_t\),
\begin{equation}
	m_d = \frac{2}{3} \mu_{tn} - \frac{1}{3} \mu_{tp}, \quad \text{and} \quad m_u = \frac{2}{3} \mu_{tp} - \frac{1}{3} \mu_{tn},
	\label{eq:13_quark_masses}
\end{equation}
where \(\mu_{t(n,p)}\) are the \(\beta\)-equilibrium values of chemical potential at \(\rho_t\).

The energy density of the non-interacting fermions, leptons and quarks, is then given by
\begin{equation}
	\mathcal{E}_i = \frac{N_c}{\pi^2} \int \limits_0^{k_{Fi}} k^2 \sqrt{m_i^2 + k^2} \dd{k},
	\label{eq:14_energy_density_noninteracting}
\end{equation}
where \(i = e,\ \mu,\ d,\ u\) and \(N_c = 3 (1)\) for quarks (leptons).
The total energy density of the system is
\begin{equation}
	\mathcal{E} = \mathcal{E}_B + \mathcal{E}_e + \mathcal{E}_\mu + \mathcal{E}_d + \mathcal{E}_u.
	\label{eq:15_total_energy_density}
\end{equation}
The total pressure is
\begin{equation}
	P = P_B + P_L + P_Q,
	\label{eq:16_total_pressure}
\end{equation}
where \(P_B\) is the hadronic pressure, \(P_L\) the lepton pressure, and \(P_Q\) the quark pressure, which are themselves given by
\begin{align}
	P_B = {} & \mu_n \rho_n + \mu_p \rho_p - \mathcal{E}_B,                       \\
	P_L = {} & \mu_e \rho_e + \mu_\mu \rho_\mu - \mathcal{E}_e - \mathcal{E}_\mu, \\
	P_Q = {} & \mu_d \rho_d + \mu_u \rho_u - \mathcal{E}_d - \mathcal{E}_u.
	\label{eq:17_19_particle_pressures}
\end{align}

\subsection{Bayesian Inference}
The quarkyonic model has seven free parameters which control the behaviour of the EoS.
Varying these parameters allows us to obtain such EoSs which satisfy observational and theoretical constraints on dense matter.
These model parameters along with the possible ranges in which they can vary (called their priors) are listed in Table~\ref{tab:priors}.
We would like to mention here that the results will be, to some extent, sensitive to our choices of priors.
While the Gaussian priors are motivated from experimentally obtained values, there is no perfect choice for the remaining parameters.
For them, we choose uniform priors as it is the simplest distribution possible.
Using a nested sampling method for exploring the parameter space is also beneficial, as each iteration progressively samples the subspace with the highest likelihood, leading to a better exploration and extraction of optimal results from the priors.
\begingroup
\squeezetable
\begin{table}[ht]
	\caption{The model parameters and their priors utilized for the Bayesian inference. These priors are either set as Gaussian priors (listed as mean $\pm$ standard deviation) for parameters which have experimentally obtained values, and uniform priors (listed as the closed range [lower limit, upper limit]) for the remaining parameters. $\rho_0 = 0.16\ \mathrm{fm}^{-3}$ is the nuclear saturation density.}\label{tab:priors}
	\begin{ruledtabular}
		\begin{tabular}{c c c}
			\textbf{Model Parameter}                                                    & \textbf{Prior type} & \textbf{Range}     \\
			\midrule
			Quarkyonic transition density ($\rho_t$) ($\mathrm{fm^{-3}}$)               & Uniform             & $\bqty{0.18, 0.8}$ \\
			QCD scale parameter $\Lambda_{qcd}$ ($\mathrm{MeV}$)~\cite{Zhao:2020dvu}    & Uniform             & $\bqty{10, 2000}$  \\
			Symmetry energy at $\rho_0$ ($J$) ($\mathrm{MeV}$)~\cite{Cartaxo:2025jpi}   & Gaussian            & $32.5 \pm 2.5 $    \\
			Slope of symmetry energy at $\rho_0$ ($L$) ($\mathrm{MeV}$)                 & Uniform             & $\bqty{20, 100}$   \\
			Incompressibility at $\rho_0$ ($K$) ($\mathrm{MeV}$)~\cite{Cartaxo:2025jpi} & Gaussian            & $230 \pm 40$       \\
			Binding energy at $\rho_0$ ($BE/A$) ($\mathrm{MeV}$)~\cite{Cartaxo:2025jpi} & Gaussian            & $16 \pm 0.2$       \\
			$\gamma_1$ term in interaction potential~\cite{Zhao:2020dvu}                & Uniform             & $\bqty{0.5, 5.0}$  \\
		\end{tabular}
	\end{ruledtabular}
\end{table}
\endgroup

Following the Bayesian inference framework developed in Ref.~\cite{Greif:2018njt, Raaijmakers:2019dks, Raaijmakers:2019qny, Raaijmakers:2021uju}, we use Bayes' theorem to calculate the posterior distribution of the model parameters, and this seven-dimensional parameter space is traversed via nested sampling using the UltraNest package~\cite{buchner_ultranest_2021}.
Obtaining the Bayes' probability for each parameter vector involves evaluating the likelihood of how well the family of NSs described by the vector agrees with multi-messenger observations and theoretical bounds (shown in~\ref{sec:likelihood}).
We obtain three separate constrained sets from the Bayesian inference process, namely $\alpha$, $\beta$ and $\gamma$.
For Set $\alpha$, we use pulsar mass and radius measurements from NICER for PSR J0030+0451~\cite{Riley:2019yda} and PSR J0740+6620~\cite{Riley:2021pdl}, the mass and tidal deformability measurements from LIGO/Virgo for the binary neutron star merger event GW170817~\cite{LIGOScientific:2018cki}, and ab-initio low-density nuclear theory bounds on EoS from $\chi$EFT calculations~\cite{Huth:2021bsp}, imposed up to the nuclear saturation density, $\rho_0=0.16\ \mathrm{fm}^{-3}$.
Set $\beta$ changes the bounds to include the recently reported NICER measurements of PSR J0437-4715~\cite{Choudhury:2024xbk} and PSR J0614-3329~\cite{Mauviard:2025dmd}, and set a maximum mass ceiling at $M_{\max} \leq 2.33\ \mathrm{M_\odot}$ using the $1\sigma$ upper bound from Ref.~\cite{Fan:2023spm}.
In Set $\gamma$, the maximum mass ceiling is reduced to $2.15\ \mathrm{M_\odot}$, using the value obtained in Ref.~\cite{Shao:2019ioq}, in order to study the consequences of stricter limits compared to Set $\beta$.
We also ensure that each EoS remains causal ($c_s^2 \leq 1$ at all densities), and produces a NS with a mass of at least $1.97\ \mathrm{M_\odot}$ to satisfy the heaviest NS mass observations~\cite{Demorest:2010bx}.
Increasing constraints from Set $\alpha$ through $\gamma$ progressively limits the final parameter spaces.

Each parameter vector from the three constrained sets ($\alpha$, $\beta$, $\gamma$) gives an EoS describing the pressure-energy density relation in NS matter, for which the Tolman–Oppenheimer–Volkoff equations~\cite{Tolman:1939jz, Oppenheimer:1939ne, Hinderer:2007mb, Hinderer:2009ca} are solved, yielding corresponding mass-radius ($M-R$) and mass-tidal deformability ($M-\Lambda$) relations.

\subsection{Likelihood Evaluation}\label{sec:likelihood}
Bayes theorem provides the posterior distribution of a vector of model parameters $\theta$, given the observational data $\mathcal{D}$, obtained using
\begin{equation}
	P(\theta | \mathcal{D}) = \frac{\mathcal{L}(\mathcal{D}|\theta) P(\theta)}{P(\mathcal{D})},
	\label{eq:20_bayes_theorem}
\end{equation}
where $P(\theta | \mathcal{D})$ is the posterior probability, $\mathcal{L}(\mathcal{D}|\theta)$ the likelihood of the data, $P(\theta)$ the prior probability of the parameters, and $P(\mathcal{D})$ the evidence.
The likelihood provides the probability of obtaining the data $\mathcal{D}$ for a particular parameter vector.
Assuming that the various data sets are statistically independent, the overall likelihood would be the product of the individual likelihoods of each data set, giving us
\begin{equation}
	\mathcal{L} = \mathcal{L}^\mathrm{PNM} \mathcal{L}^\mathrm{GW} \mathcal{L}^\mathrm{NICER}.
	\label{eq:21_total_likelihood}
\end{equation}
The individual likelihoods are computed as follows,
\begin{enumerate}
	\item $\mathcal{L}^\mathrm{PNM}$: We use pure neutron matter (PNM) energy-per-neutron constraints from several $\chi$EFT calculations presented in~\cite{Huth:2021bsp} at densities of 0.08, 0.12, and 0.16 fm$^{-3}$. We adopt a Gaussian likelihood, which reads
	      \begin{equation}
		      \mathcal{L}^\mathrm{PNM}(\bm{\theta}) = \prod_{j=1}^{N_\mathrm{PNM}} \frac{1}{\sqrt{2\pi\ \sigma_j^2}} \exp\bqty{-\frac{1}{2} \pqty{\frac{D_j - m_j(\bm{\theta})}{\sigma_j}}^2},
		      \label{eq:22_PNM_likelihood}
	      \end{equation}
	      where the index $j$ runs over the $N_\mathrm{PNM} = 3$ density points ($0.08$, $0.12$, and $0.16\ \mathrm{fm}^{-3}$), with $D_j$ denoting the energy per neutron from the $\chi$EFT calculation at density $\rho_j$, $m_j(\bm{\theta})$ the corresponding model prediction, and $\sigma_j$ the associated uncertainty from~\cite{Huth:2021bsp}.
	\item $\mathcal{L}^\mathrm{GW}$: We employ the tidal deformability posterior samples for GW170817~\cite{LIGOScientific:2018hze} within our inference framework, using a Kernel Density Estimation (KDE) to construct the probability distribution $P(\mathcal{D}_{\rm GW} | m_1, m_2, \Lambda_1, \Lambda_2)$ from the masses ($m_1, m_2$) and corresponding tidal deformabilities ($\Lambda_1, \Lambda_2$) of the two binary components. The likelihood is given by~\cite{Imam:2024gfh},
	      \begin{align}
		      \mathcal{L}^\mathrm{GW} = & \int_{m_2}^{M_u} \dd{m_1} \int_{M_l}^{m_1} \dd{m_2}\ P\pqty{m_1 | \theta}P\pqty{m_2 | \theta}         \\
		                                & \cp P \pqty{\mathcal{D}_{GW} | m_1, m_2, \Lambda_1 (m_1, \theta), \Lambda_2 (m_2, \theta)}, \nonumber
		      \label{eq:23_GW_likelihood}
	      \end{align}
	      where, $P(m|\theta)$ is given by,
	      \begin{equation}
		      P (m | \theta) =
		      \begin{cases}
			      \frac{1}{M_u - M_l}, & \text{iff}\ M_l \leq m \leq M_u, \\
			      0,                   & \text{otherwise},
		      \end{cases}
		      \label{eq:24_probability_m_given_theta_GW}
	      \end{equation}
	      with $M_l = 1 \ \mathrm{M}_\odot$ and $M_u = M_{\max}(\theta)$~\cite{Malik:2024qjw}.
	\item $\mathcal{L}^\mathrm{NICER}$: For the parameter vector to align with the NICER measurements of NS mass and radius, the likelihood is evaluated as~\cite{Imam:2024gfh, Cartaxo:2025jpi}
	      \begin{equation}
		      \mathcal{L}^\mathrm{NICER} = \int_{M_l}^{M_u} dm\ P(m | \theta)\ P(\mathcal{D}_\mathrm{NICER} | m, R(m, \theta)).
		      \label{eq:L_NICER}
	      \end{equation}
	      Here, $P(m | \theta)$ is the same as Eq.~\eqref{eq:24_probability_m_given_theta_GW}.
	      This is evaluated for the pulsars PSR J0030+0451~\cite{Riley:2019yda}, PSR J0740+6620~\cite{Riley:2021pdl}, PSR J0437-4715~\cite{Choudhury:2024xbk}, and PSR J0614-3329~\cite{Mauviard:2025dmd} separately and added in the different combinations for the three result sets.
\end{enumerate}

\section{Results}
\begin{figure*}
	\includegraphics[width=\linewidth]{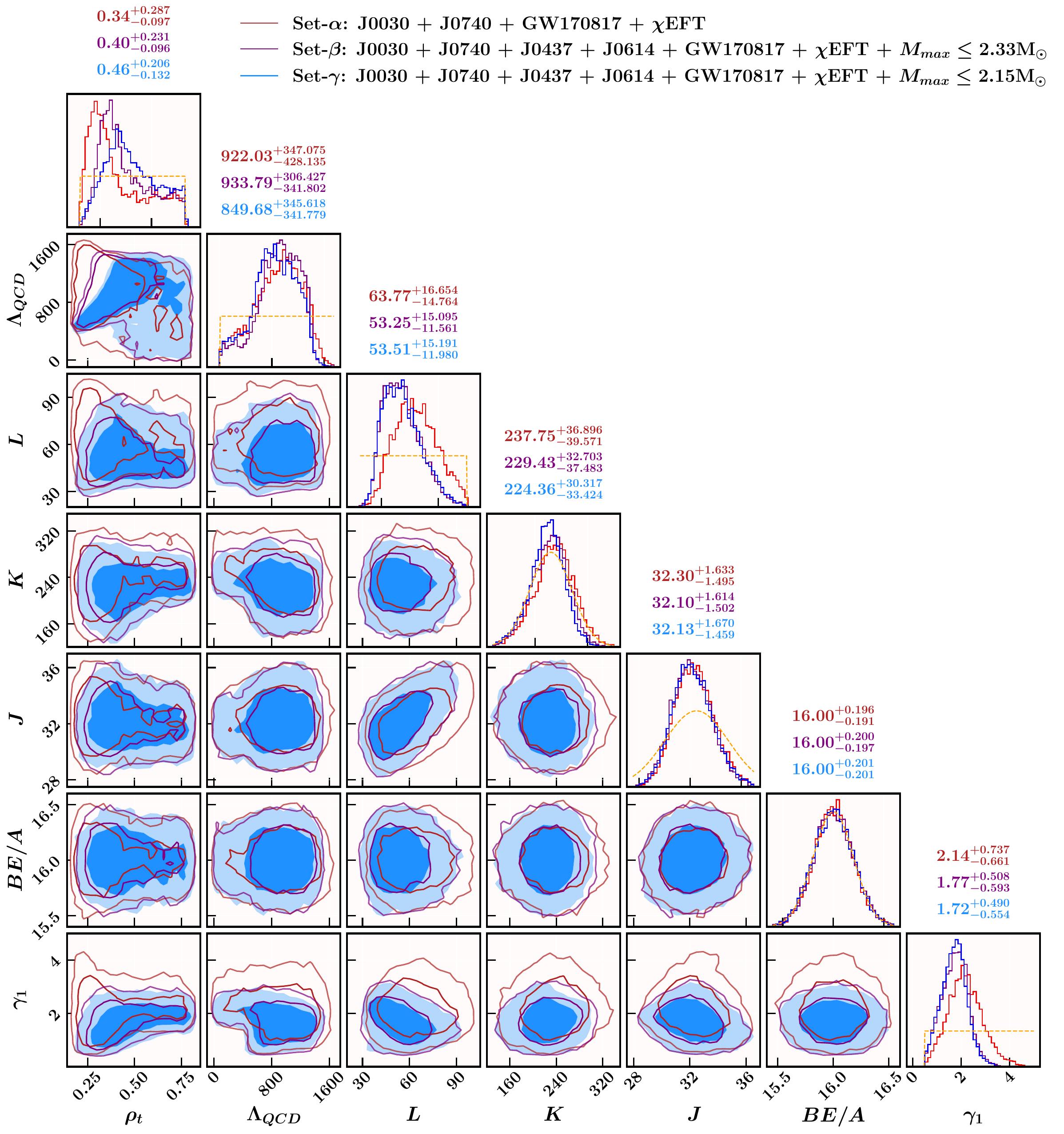}
	\caption{Correlation corner-plot of the seven parameters constrained via Bayesian inference at $1$, and $2\sigma$ confidence levels. Posterior distribution histogram alongside constrained values of parameters are shown along the diagonal. The prior distributions of the parameters are shown using the dashed orange line in the histograms. Red, purple, and blue are used for sets $\alpha$, $\beta$ and $\gamma$, respectively.}\label{fig:1_param_cornerplot}
\end{figure*}

Bayesian inference of the quarkyonic matter model's seven free parameters allows us to utilize available constraints to identify the values (or ranges of values) the parameters must take to yield EoSs that satisfy the constraints placed on them.
Table~\ref{tab:bayesian_count} shows the final number of valid vectors of parameters from each constrained set obtained from the Bayesian inference process.
In Figure~\ref{fig:1_param_cornerplot}, we present the results of the inference process in the form of a correlation corner-plot, with the three constrained sets \(\alpha,\ \beta\), and \(\gamma\) being depicted in red, purple and blue, respectively.
The correlations plots are shown using two contours with the darker (lighter) region marking the \(1\sigma\) (\(2\sigma\)) boundary.
The posterior distributions of the parameters are shown as histograms along the diagonal, with the orange dashed line showing the priors taken for them.
Above the diagonal, we also provide the median values along with the differences to the upper and lower \(1\sigma\) bounds.

\begingroup
\squeezetable
\begin{table}[ht]
	\caption{The constraints imposed and the final number of valid parameter vectors (i.e. the total number of valid EoSs) obtained from the Bayesian inference process for each result set. Here, the pulsars are denoted in short using J0030 for PSR J0030+0451, J0740 for PSR J0740+6620, J0437 for PSR J0437-4715 and J0614 for PSR J0614-3329. Since the constraints imposed on Set-$\alpha$ is common to all sets, we denote this combination of constraints collectively as $C_\alpha$.}\label{tab:bayesian_count}
	\begin{ruledtabular}
		\begin{tabular}{c c c}
			\textbf{Set} & \textbf{Constraints}                                               & \textbf{Total EoSs} \\
			\midrule
			$\alpha$     & J0030 + J0740 + GW170817 + $\chi$EFT                               & $17,544$            \\
			$\beta$      & $C_\alpha$ + J0437 + J0614 + $M_{\max} \leq 2.33 \mathrm{M_\odot}$ & $18,007$            \\
			$\gamma$     & $C_\alpha$ + J0437 + J0614 + $M_{\max} \leq 2.15 \mathrm{M_\odot}$ & $19,328$
		\end{tabular}
	\end{ruledtabular}
\end{table}
\endgroup

The transition density from hadronic to quarkyonic matter, \(\rho_t\), which had a uniform prior shows a posterior distribution with a Gaussian-like peak at lower density which then becomes uniform at higher density.
The Gaussian region's peak shifts to a higher density as we go from the less restrictive Set-\(\alpha\) to the most restrictive Set-\(\gamma\), which is accompanied by a broadening of its width.
Physically, lowering the maximum allowed mass limit makes it possible for the stars in sets $\beta$ and $\gamma$ to have relatively lower inner core pressures than Set-$\alpha$.
This means that an early appearance of quarks is not required, increasing the density at which they appear and leading to the increase in \(\rho_t\).
The flattening of the distribution at high transition densities indicates that models with $\rho_t$ extending up to the prior upper bound continue satisfying all current astronomical constraints, making existing observations insufficient for fully constraining $\rho_t$.

The \(\Lambda_\mathrm{QCD}\) posteriors also show Gaussian-like distributions resulting from the inference process, with a slight flattening observed at the lower limit.
While the locations of the medians do change between sets \(\alpha,\ \beta\) and \(\gamma\), a general agreement can be seen between the three histograms indicating that the additional constraints in sets \(\beta\) and \(\gamma\) do not have a significant contribution towards constraining \(\Lambda_\mathrm{QCD}\).
The symmetry energy slope \(L\) posterior also forms a Gaussian, and the imposition of the additional constraints in sets \(\beta\) and \(\gamma\) causes the Gaussian to narrow while lowering the location of the median.
Similar behaviour is also demonstrated by the \(\gamma_1\) parameter.

Of the parameters with Gaussian priors, we do not see any significant impact of the Bayesian inference process for the incompressibility and the binding energy, with their posteriors and priors not having any noticeable difference.
The symmetry energy \(J\) have similar Gaussian-like posteriors, and are all narrower than the Gaussian prior.
While their overlap indicates that the additional astronomical constraints in sets \(\beta\) and \(\gamma\) do not have much of a contribution, the narrowing of the distributions coupled with their lower medians reflects the effectiveness of combining astronomical observations with terrestrial nuclear experiments to give a better understanding of dense matter.

\begin{figure}
	\centering
	\includegraphics[width=\columnwidth]{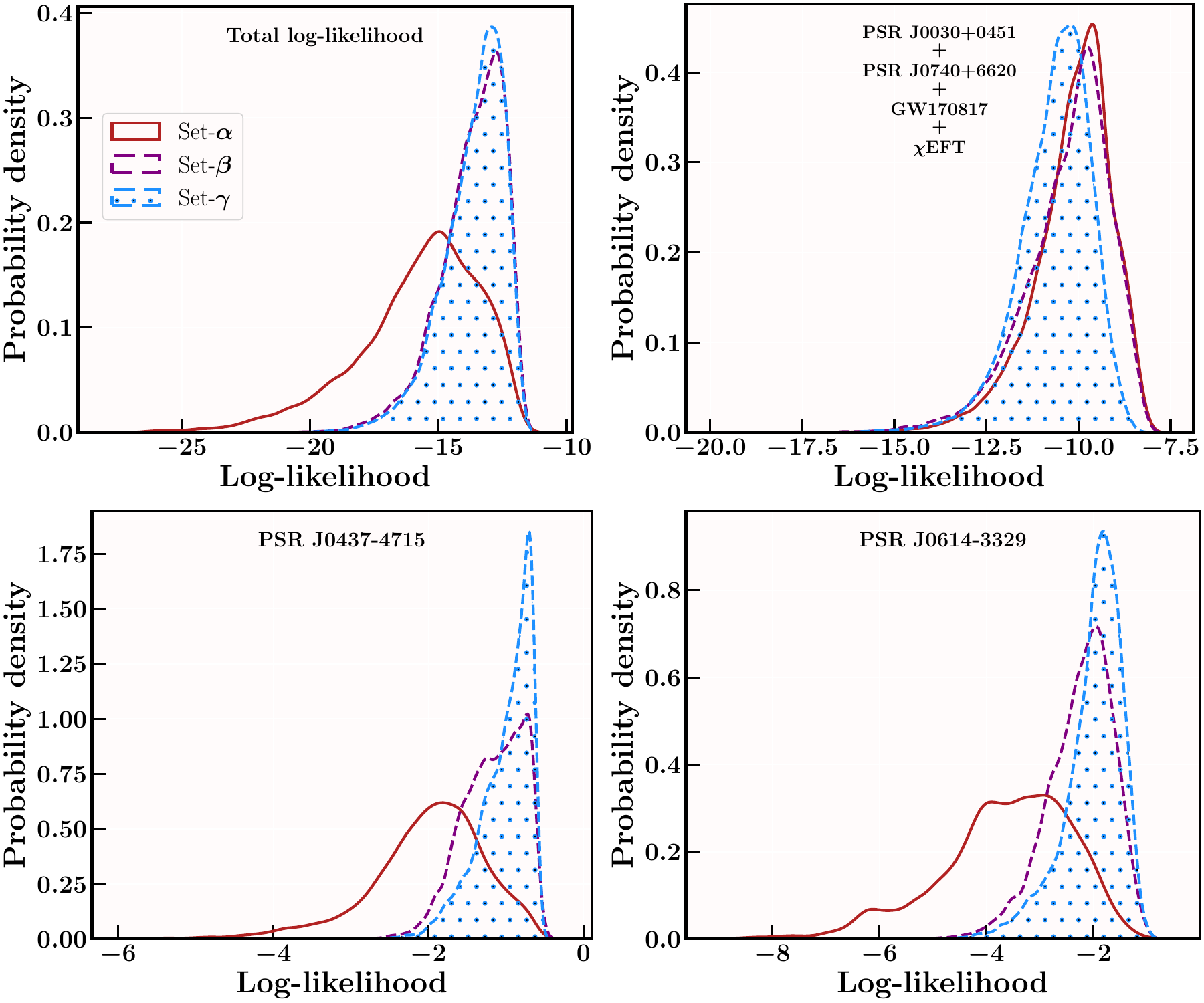}
	\caption{Probability density distributions of the log-likelihoods of the \(\alpha,\ \beta\) and \(\gamma\) constrained sets shown using red, purple and blue, respectively. The sub-figures display the distributions when log-likelihoods are calculated using all available datasets (upper-left), common datasets for all result sets (upper-right), only PSR J0437-4715 (lower-left), and only PSR J0614-3329 (lower-right).}\label{fig:2_log-likelihood}
\end{figure}

\begin{figure*}
	\subfloat[]{\includegraphics[width=0.33\textwidth, keepaspectratio]{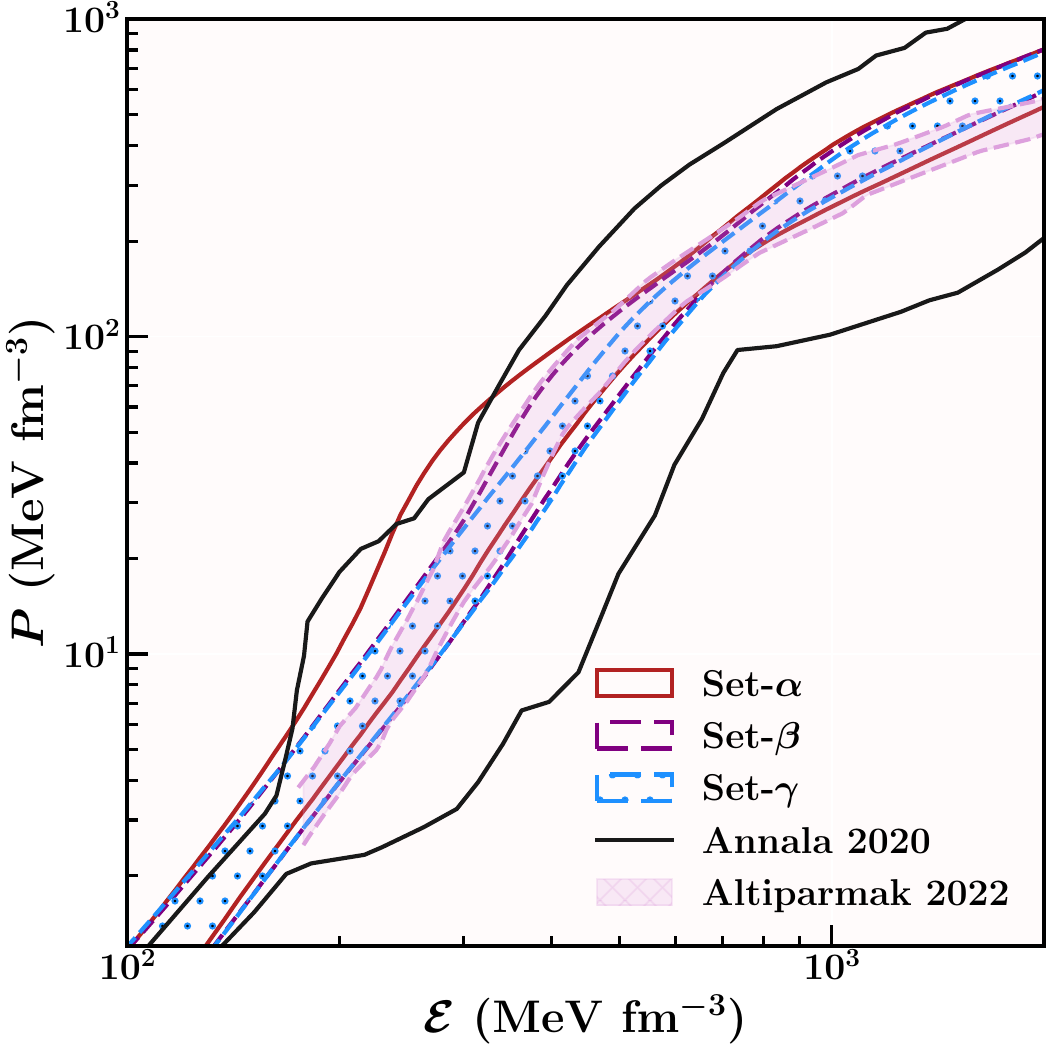}\label{fig:3a_EoS}}
	\subfloat[]{\includegraphics[width=0.33\textwidth, keepaspectratio]{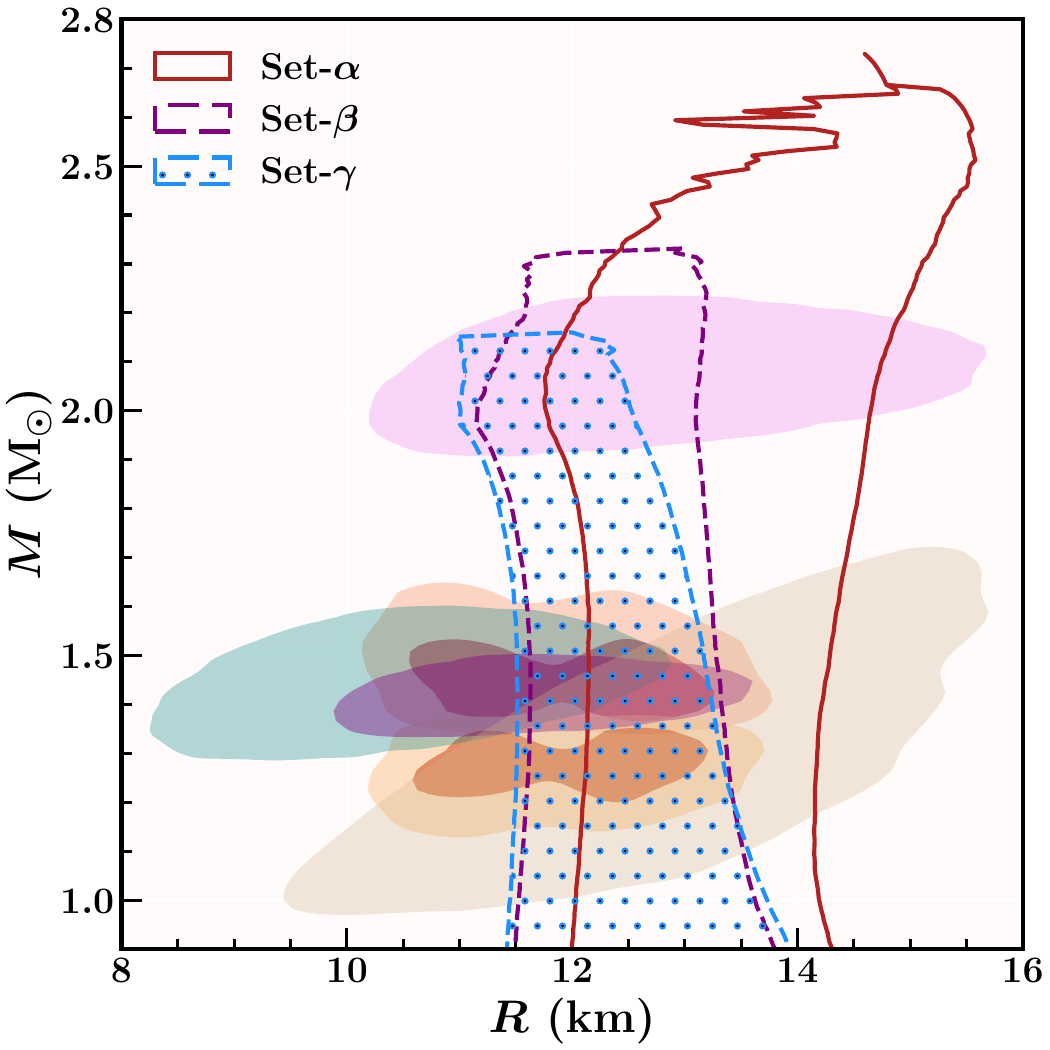}\label{fig:3b_MR}}
	\subfloat[]{\includegraphics[width=0.33\textwidth, keepaspectratio]{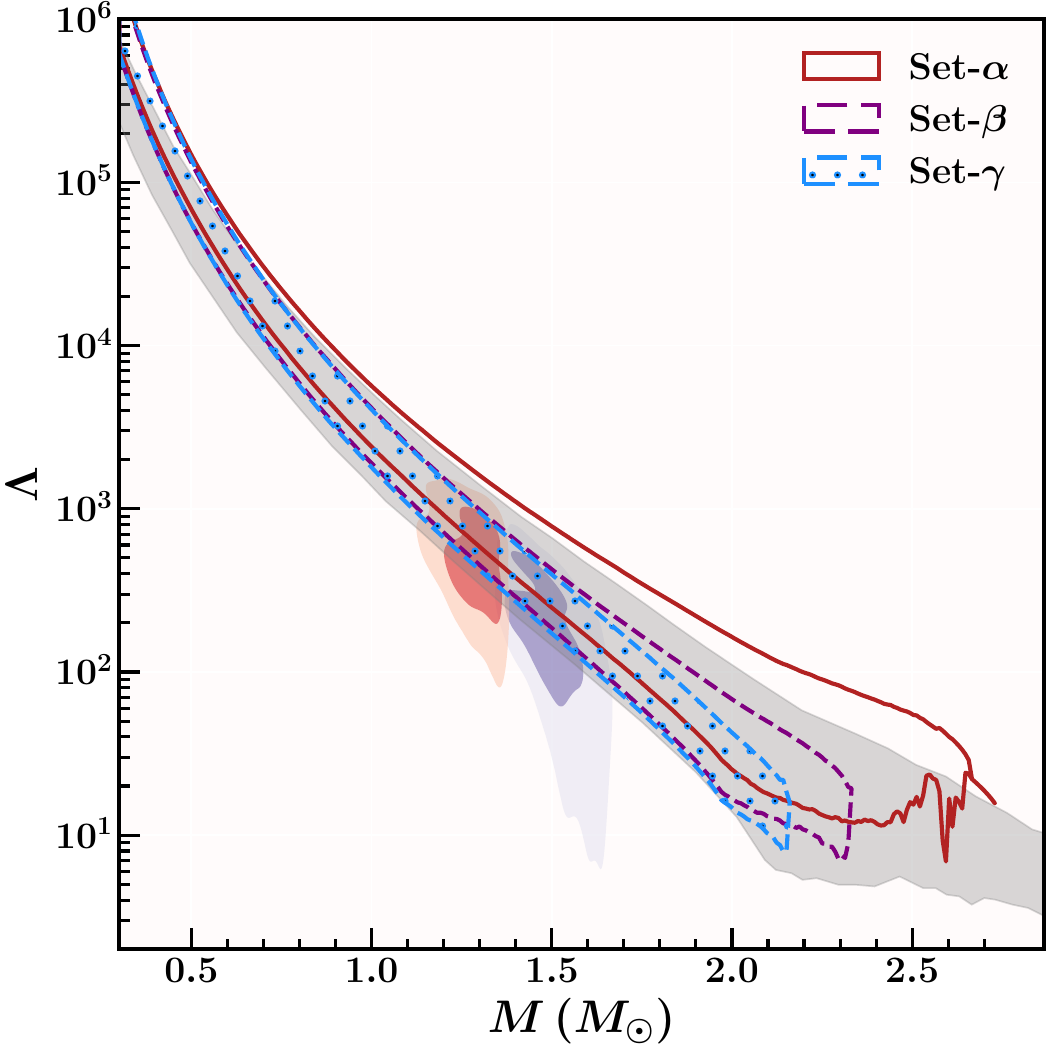}\label{fig:3c_ML}}
	\caption{1-dimensional posteriors at $90\%$ confidence interval of, (a) pressure on energy density grid, (b) radius on mass grid, and (c) dimensionless tidal deformability on mass grid, for sets $\alpha$, $\beta$, and $\gamma$ - represented in solid red boundary, dashed purple boundary, and dashed blue boundary (with blue dots inside), respectively. Bounds on EoS from Refs.~\cite{Annala:2019puf, Altiparmak:2022bke} are shown in (a) using solid black lines and the salmon shaded region. Colored patches in (b) show the $2\sigma$ bounds from NICER observations of PSR J0030+0451 (tan)~\cite{Riley:2019yda}, PSR J0740+6620 (salmon)~\cite{Riley:2021pdl}, PSR J0437-4715 (teal)~\cite{Choudhury:2024xbk} and PSR J0614-3329 (violet)~\cite{Mauviard:2025dmd}, and the $90\%$ ($50\%$) bounds from GW170817 in light (dark) orange~\cite{LIGOScientific:2018cki}. The gray region is (c) is the $90\%$ confidence region the $M-\Lambda$ posterior from GW170817~\cite{noauthor_ligo-p1800115-v12_nodate}, and the observed $90\%$ ($50\%$) values for GW170817 components are shown using the light (dark) red and purple regions~\cite{LIGOScientific:2018cki}.}\label{fig:2_macro_posterior}
\end{figure*}

\begin{figure*}
	\centering
	\includegraphics[width=\textwidth, keepaspectratio]{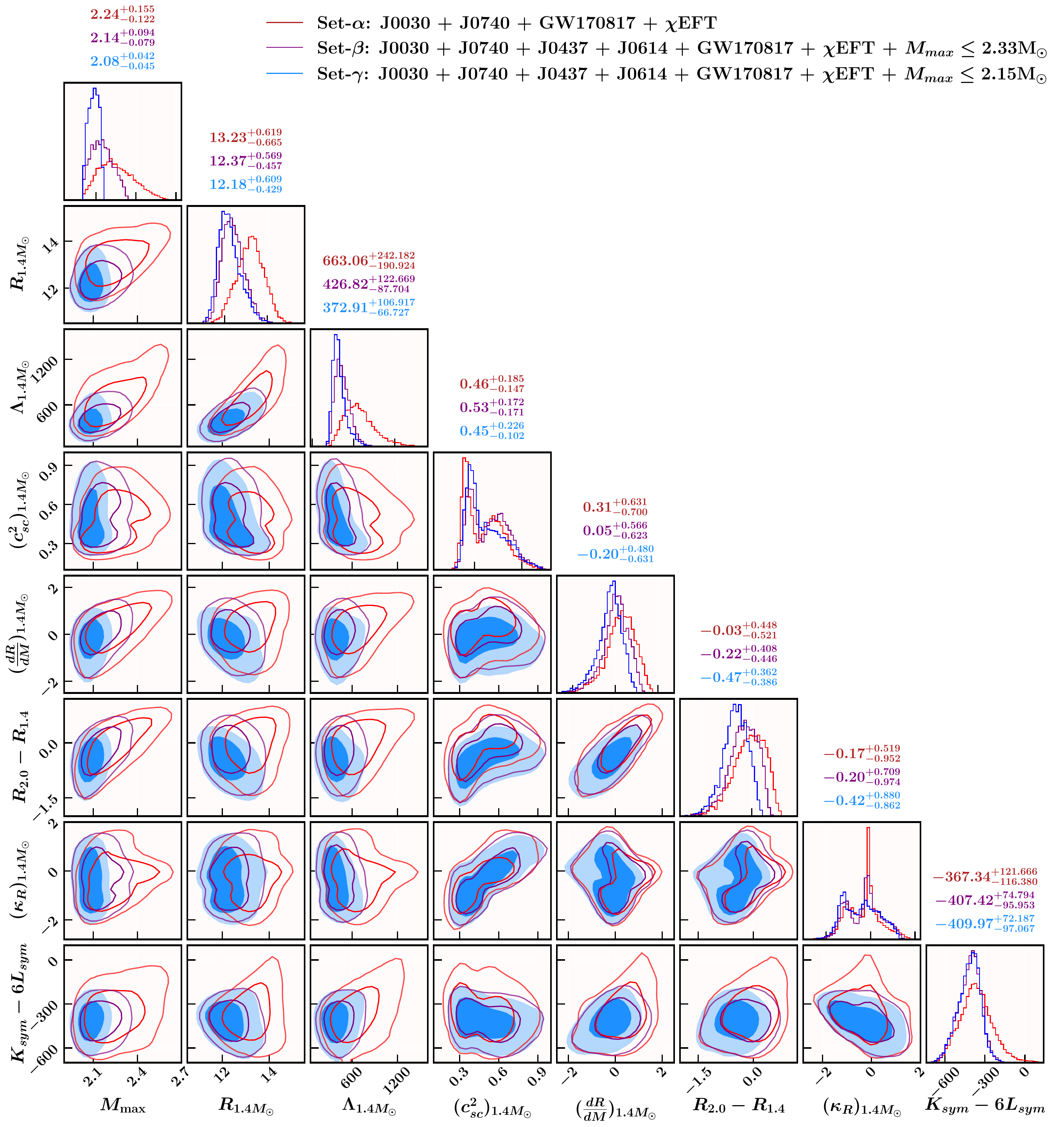}
	\caption{Corner-plot showing posterior distributions and correlations between select NS and nuclear matter properties for the three constrained sets. The off-diagonal plots show the results at $1$ and $2\sigma$ bounds, while the diagonal plots show the total distribution of the quantities. The sets $\alpha$, $\beta$ and $\gamma$ are represented using red, purple and blue color, respectively.}\label{fig:4_posterior_cornerplot}
\end{figure*}

Fig.~\ref{fig:2_log-likelihood} shows the results of the Bayesian inference in terms of how well the parameter vectors in the constrained results match with the constraints used.
The match is quantified using the log-likelihood values, and we use the value calculated for different combinations of constraints to better understand it.
The upper-left figure shows the distribution of total log-likelihood, i.e., when all constraints (PNM + GW170817 + \(\chi\)EFT + all NICER) are used to calculate the value.
On the upper-right we have the log-likelihood for the constraints imposed commonly to all the three sets, PNM + GW170817 + \(\chi\)EFT + PSR J0030+0451 + PSR J0740+6620.
In the lower-half, we have log-likelihoods calculated for individual pulsars, with PSR J0437-4715 on the left and PSR J0614-3329 on the right.
The upper-left figure shows Set-\(\alpha\) having a lower overall agreement with the constraints as compared to \(\beta\) or \(\gamma\), that can be seen (from the two figures on the lower-half) to arise from its low agreement with the pulsars PSR J0437-4715 and PSR J0614-3329.
While Set-\(\alpha\) was indeed not generated by considering the two pulsars as constraints, being subject to only the common constraints causes it to have a bias towards high masses and large radii (Fig.~\ref{fig:3b_MR} and Fig.~\ref{fig:4_posterior_cornerplot}).
The upper-right figure shows Set-$\beta$ having a distribution similar to Set-$\alpha$, but has higher likelihoods for the individual pulsars (lower-half of Fig.~\ref{fig:2_log-likelihood}) leading to a better total likelihood distribution.
Set-$\gamma$ has a similar total log-likelihood distribution to $\beta$, indicating that more high-mass NS measurements are required to better estimate the NS $M_{\max}$ limit.

The outcome of the Bayesian inference for the EoS and astronomical observables is displayed in Fig.~\ref{fig:2_macro_posterior}, where we show the posterior distributions of the NS pressure-energy density, mass-radius and mass-tidal deformability relations.
The EoS plot in Fig.~\ref{fig:3a_EoS} shows the 1-dimensional posterior distribution of pressure ($P$) on a grid of energy density ($\mathcal{E}$).
The $90\%$ confidence intervals of the pressure posteriors of sets $\alpha$, $\beta$ and $\gamma$ are shown by the regions bounded by solid red, dashed purple, and dashed blue lines, respectively.
\citeauthor{Annala:2019puf} had utilized a speed of sound interpolation method to generate EoSs satisfying the $M\geq 1.97\ \mathrm{M_\odot}$ and $70 < \Lambda_{1.4} < 580$~\cite{LIGOScientific:2018cki} constraints, and selecting the causally allowed ($c_s^2 \leq 1$) EoSs gives a bound on the $P-\mathcal{E}$ space denoted by the solid black lines~\cite{Annala:2019puf}.
The region where EoSs are most likely to lie, as identified by~\citeauthor{Altiparmak:2022bke} using a similar approach as Ref.~\cite{Annala:2019puf}, is shown in salmon color~\cite{Altiparmak:2022bke}.
Fig.~\ref{fig:3b_MR} shows the 1-dimensional posterior distribution of the stellar radius ($R$) on a grid of the stellar mass ($M$) using the same color scheme as Fig.~\ref{fig:3a_EoS}.
The colored patches correspond to astronomical observations, showing the $2\sigma$ NICER measurements of pulsars PSR J0030+0451 (tan)~\cite{Riley:2019yda}, PSR J0740+6620 (salmon)~\cite{Riley:2021pdl}, PSR J0437-4715 (teal)~\cite{Choudhury:2024xbk} and PSR J0614-3329 (violet)~\cite{Mauviard:2025dmd}, and the $90\%$ ($50\%$) confidence regions of GW170817's components (light and dark orange)~\cite{LIGOScientific:2018cki}.
Fig.~\ref{fig:3c_ML} uses a mass ($M$) grid to show the 1-dimensional posterior distribution of the dimensionless tidal deformability ($\Lambda$) calculated using the formalism in Ref.~\cite{Hinderer:2007mb, Hinderer:2009ca}.
The measured values of the lighter and heavier components in GW170817 are denoted by the red and purple patches, respectively, with the $90\%$ and $50\%$ confidence regions marked with a lighter and darker shade.
The gray shaded region is the $90\%$ credible level of the full $M-\Lambda$ posterior from GW170817 obtained from parametric EoSs~\cite{LIGOScientific:2018hze, noauthor_ligo-p1800115-v12_nodate}.

As the limitations on the mass, radius and tidal deformability are increased going from Set-$\alpha$ to $\gamma$, the posterior distributions shrink, leading to better fits with the observations.
There is a noticeable hump in Set-$\alpha$'s pressure posterior in the $\mathcal{E} \sim 200-400\ \mathrm{MeV\ fm^{-3}}$ region that extends beyond the~\citeauthor{Annala:2019puf} bounds in black, something not seen in sets $\beta$ and $\gamma$, and this sharp stiffening of the EoSs is directly related to them yielding high masses and large radii.
Unlike our constraints,~\citeauthor{Annala:2019puf} had strict bounds for $\Lambda_{1.4}$, and with no small radii observations to motivate more compact NSs, Set-$\alpha$ yields wide $P$, $R$ and $\Lambda_{1.4}$ posteriors.
The two low radii observations near canonical mass that sets $\beta$ and $\gamma$ need to satisfy is analogous to the tight $\Lambda_{1.4}$ constraints of Ref.~\cite{Annala:2019puf}, leading to them not having a sharp pressure increase in the intermediate energy-density region.

The impact of imposing the low radii constraints from PSR J0437-4715 and PSR J0614-3329 is clearly evident in the posterior distributions of \(R_{1.4\ \mathrm{M_\odot}}\) and \(\Lambda_{1.4\ \mathrm{M_\odot}}\) of sets \(\beta\) and \(\gamma\), in Fig.~\ref{fig:4_posterior_cornerplot}.
In Fig.~\ref{fig:4_posterior_cornerplot}, we have a corner-plot where we show the correlations between different quantities of interest on the off-diagonal sub-plots and along the diagonal we show their posterior distribution histograms.
Similar to Fig.~\ref{fig:1_param_cornerplot}, the medians and \(1\sigma\) bounds are presented above the diagonal, and the same colour scheme is in use.
The first quantity of interest is the maximum mass \(M_{\max}\) obtained for each parameter vector.
We also wish to explore how NS and nuclear properties might be correlated, and if the behaviour is impacted by the constraints imposed.
For this we take the leading isospin incompressibility term, \(K_\tau \sim K_{\mathrm{sym}} - 6 L_{\mathrm{sym}}\).
Since our constraints are all near the canonical mass limit of \(1.4\ \mathrm{M_\odot}\), and considering that most observed pulsars also lie in the vicinity of this region, we study our other quantities of interest at the canonical limit.
These quantities are radius \(R_{1.4\mathrm{M_\odot}}\), dimensionless tidal deformability \(\Lambda_{1.4\mathrm{M_\odot}}\), sound speed at the core of the canonical NS \(\pqty{c_{sc}^2}_{1.4\mathrm{M_\odot}}\), the slope of the \(M-R\) relation \(\pqty{\dv{R}{M}}_{1.4\mathrm{M_\odot}}\), the curvature of the \(M-R\) relation \(\pqty{\kappa_R}_{1.4\mathrm{M_\odot}}\), and the difference in radii of a \(2\ \mathrm{M_\odot}\) and a \(1.4\ \mathrm{M_\odot}\) NS, \(\Delta R = R_{2.0} - R_{1.4}\).
The curvature \(\kappa_R\) of the \(R(M)\) relation is expressed as~\cite{Bauswein:2025dfg}
\begin{equation}
	\kappa_R = \frac{\dv*[2]{R}{M}}{\pqty{1 + \pqty{\dv*{R}{M}}^2}^{3/2}}.
	\label{eq:26_curvature}
\end{equation}

\begingroup
\squeezetable
\begin{table*}[tp!]
	\caption{Median and \(2\sigma\) confidence limits for neutron star and nuclear matter properties from the three constrained sets.}\label{tab:posteriors}
	\begin{center}
		\begin{ruledtabular}
			\begin{tabular}{c c c c c c c c c c}
				                                                                     & \multicolumn{3}{c}{\textbf{Set-$\alpha$}} & \multicolumn{3}{c}{\textbf{Set-$\beta$}} & \multicolumn{3}{c}{\textbf{Set-$\gamma$}}                                                                                                                                                 \\
				\cline{2-4} \cline{5-7} \cline{8-10}
				\textbf{Quantity}                                                    & \textbf{Median}                           & \textbf{$2\sigma$ Upper}                 & \textbf{$2\sigma$ Lower}                  & \textbf{Median} & \textbf{$2\sigma$ Upper} & \textbf{$2\sigma$ Lower} & \textbf{Median} & \textbf{$2\sigma$ Upper} & \textbf{$2\sigma$ Lower} \\
				\midrule
				$\rho_t~(\mathrm{fm^{-3}})$                                          & $0.3448$                                  & $1.7483$                                 & $0.2105$                                  & $0.4046$        & $0.7501$                 & $0.2607$                 & $0.4636$        & $0.7634$                 & $0.2608$                 \\
				$\Lambda_{QCD}~(\mathrm{MeV})$                                       & $922.0348$                                & $1444.0031$                              & $182.6728$                                & $934.7894$      & $1364.9343$              & $225.0693$               & $849.6760$      & $1350.4611$              & $190.3005$               \\
				$L~(\mathrm{MeV})$                                                   & $63.7687$                                 & $91.2295$                                & $41.3011$                                 & $53.2478$       & $79.1214$                & $36.4845$                & $53.5054$       & $80.3671$                & $36.5260$                \\
				$K~(\mathrm{MeV})$                                                   & $237.7519$                                & $299.6299$                               & $168.2071$                                & $229.4321$      & $283.2684$               & $164.3698$               & $224.3606$      & $274.9880$               & $162.3285$               \\
				$J~(\mathrm{MeV})$                                                   & $32.2998$                                 & $35.1199$                                & $29.8532$                                 & $ 32.1001$      & $34.8990$                & $29.6387$                & $32.1267$       & $35.0145$                & $29.6977$                \\
				$BE/A~(\mathrm{MeV})$                                                & $16.0027$                                 & $16.3453$                                & $15.6695$                                 & $16.0026$       & $16.3421$                & $15.6708$                & $16.0022$       & $16.3413$                & $15.6624$                \\
				$\gamma_1$                                                           & $2.1450$                                  & $3.4787$                                 & $1.0297$                                  & $1.7694$        & $2.6681$                 & $0.8370$                 & $1.7248$        & $2.6002$                 & $0.8314$                 \\
				\midrule
				$M_{\max}~(\mathrm{M}_\odot)$                                        & $2.2359$                                  & $2.4967$                                 & $2.0409$                                  & $2.1317$        & $2.2851$                 & $1.9982$                 & $2.0726$        & $2.1393$                 & $1.9871$                 \\
				$R_{1.4\ \mathrm{M_\odot}}~(\mathrm{km})$                            &
				$13.2260$                                                            & $14.2436$                                 & $12.1146$                                & $12.3560$                                 & $13.3540$       & $11.6107$                & $12.1816$                & $13.2660$       & $11.4966$
				\\
				$\Lambda_{1.4\ \mathrm{M_\odot}}$                                    & $661.2378$                                & $1130.3473$                              & $373.5515$                                & $423.4336$      & $649.9620$               & $286.2540$               & $372.9509$      & $617.6850$               & $267.1324$               \\
				$\pqty{c_{sc}^2}_{1.4\mathrm{M_\odot}}$                              &
				$0.4586$                                                             & $0.7649$                                  & $0.2836$                                 & $0.5080$                                  & $0.7986$        & $0.3070$                 & $0.4274$                 & $0.7808$        & $0.3018$
				\\
				$\pqty{\dv*{R}{M}}_{1.4\mathrm{M_\odot}}~(\mathrm{km\ M_\odot^{-1}})$ & $0.2953$                                  & $1.2905$                                 & $-1.1061$                                 & $0.0136$        & $0.9204$                 & $-1.2421$                & $-0.2307$       & $0.5638$                 & $-1.4927$                \\
				$R_{2} - R_{1.4}~(\mathrm{km})$                                      & $-0.0337$                                 & $0.6165$                                 & $-0.9773$                                 & $-0.2182$       & $0.3642$                 & $-1.0488$                & $-0.4743$       & $0.0783$                 & $-1.1987$                \\
				$\pqty{\kappa_R}_{1.4M_\odot}\ \mathrm{(M_\odot\ km^{-2})}$          & $-0.1745$                                 & $1.0660$                                 & $-1.5451$                                 & $-0.2273$       & $0.9861$                 & $-1.5487$                & $-0.5120$       & $0.8969$                 & $-1.6505$                \\
				$K_{sym} - 6 L_{sym}~(\mathrm{MeV})$                                 & $-367.5260$                               & $-142.2580$                              & $-557.8370$                               & $-405.8660$     & $-279.2334$              & $-567.7539$              & $-408.3280$     & $-283.0873$              & $-571.1020$              \\
			\end{tabular}
		\end{ruledtabular}
	\end{center}
\end{table*}
\endgroup
The impacts from the additional constraints in sets \(\beta\) and \(\gamma\) also manifest themselves in the \(\pqty{\dv{R}{M}}_{1.4\mathrm{M_\odot}}\), \(\pqty{\kappa_R}_{1.4\mathrm{M_\odot}}\), and \(\Delta R = R_{2.0} - R_{1.4}\) posteriors.
The most significant result, however, visible from the posterior distributions is the formation of bimodal distributions for \(c_{sc}^2\) and \(\kappa_R\) which we shall explore further.
For the correlations, we look for those which remain intact across the three sets.
This gives us the known positive correlations between \(R_{1.4\mathrm{M_\odot}}\) and \(\Lambda_{1.4\mathrm{M_\odot}}\), and the one between \(\pqty{\dv{R}{M}}_{1.4 \mathrm{M_\odot}}\) and \(\Delta R\)~\cite{Tang:2025xib}.
For \(\pqty{c_{sc}^2}_{1.4 \mathrm{M_\odot}}\), we discover a strong positive correlation with \(\pqty{\kappa_R}_{1.4 \mathrm{M_\odot}}\), along with weaker positive correlations with \(\pqty{\dv{R}{M}}_{1.4 \mathrm{M_\odot}}\) and \(\Delta R\), together with
a weak negative correlation with \(K_\tau\).
The quantity \(K_\tau\) has a stronger negative correlation with \(\pqty{\kappa_R}_{1.4 \mathrm{M_\odot}}\).
Table~\ref{tab:posteriors} contains the median and $2\sigma$ limit values of the quantities shown in Figs.~\ref{fig:1_param_cornerplot} and~\ref{fig:4_posterior_cornerplot}.

The NS \(M-R\) relation has a one-to-one relation with the EoS, and the slope maps the EoS's stiffness~\cite{Ferreira:2025dat}.
This has caused the slope and curvature to emerge as promising candidates for inferring the internal makeup of NS, with the mapping being used to infer the presence of hyperons~\cite{Ferreira:2025dat, Bauswein:2025dfg} or a Maxwell-type first-order phase transition to quark matter~\cite{Tang:2025xib} in NS interiors.
Observational estimates of $\dv*{M}{R}$ at specific masses have also been shown to potentially constrain nuclear matter properties~\cite{Ferreira:2025dat, Tang:2025xib}, as indicated by the correlation we observed from Fig.~\ref{fig:4_posterior_cornerplot}.
Here we follow Ref.~\cite{Bauswein:2025dfg}'s example and use \(\dv*{R}{M}\) as it prevents encountering infinities which are otherwise found in the \(M(R)\) relation's slope.

\begin{figure}
	\centering
	\includegraphics[width=\linewidth, keepaspectratio]{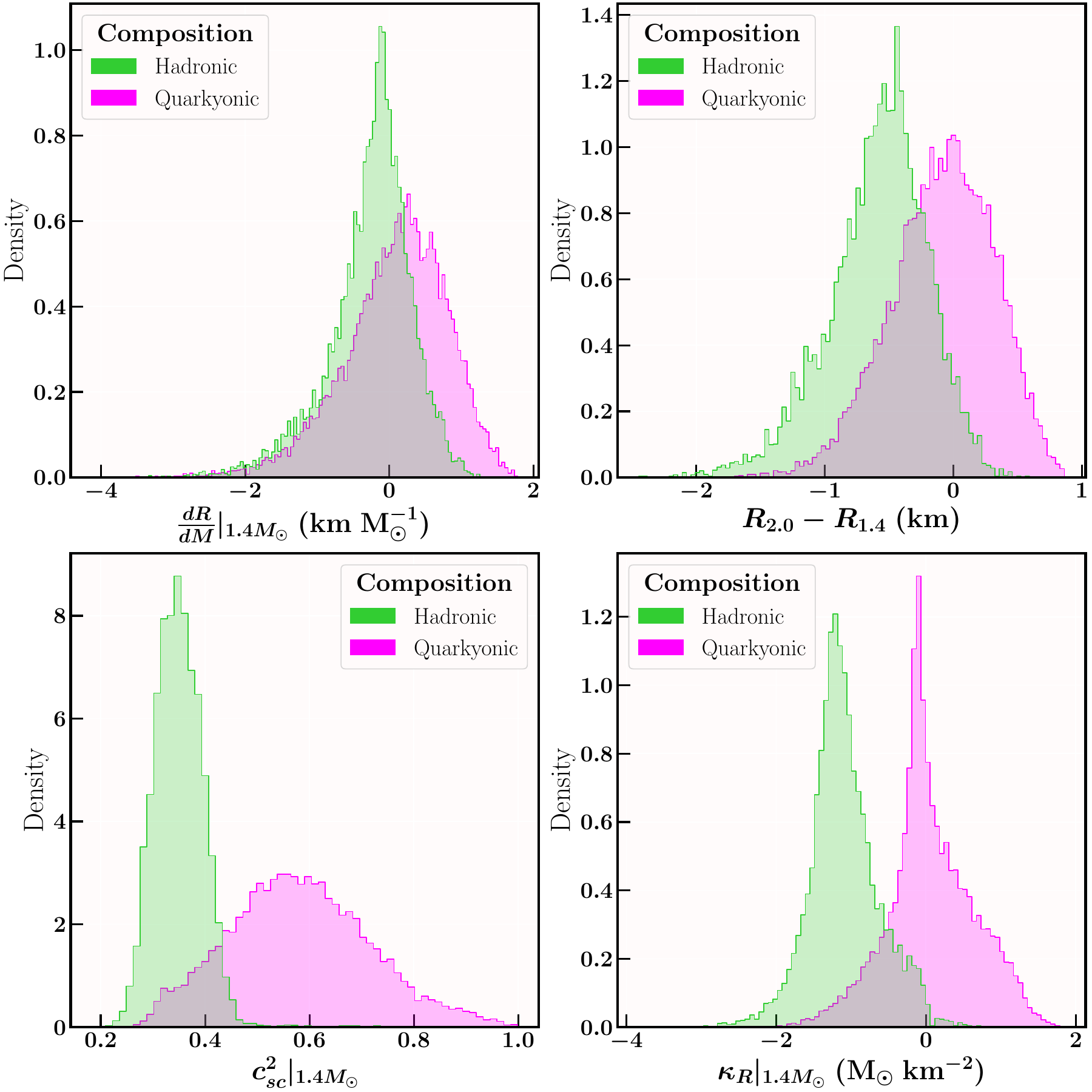}
	\caption{Marginalized posterior distributions of \(\pqty{\dv*{R}{M}}_{1.4 \mathrm{M_\odot}}\), \(\Delta R\), \(\pqty{c_{sc}^2}_{1.4 \mathrm{M_\odot}}\) and \(\pqty{\kappa_R}_{1.4 \mathrm{M_\odot}}\) on the basis of presence/absence of a quarkyonic phase in the canonical NS\@. The quarkyonic NSs are depicted using magenta, while the absence of quarkyonic phase -hadronic NSs- is shown in green.}\label{fig:5_distribution_histogram}
\end{figure}

To understand the cause of bimodality seen in \(\pqty{c_{sc}^2}_{1.4 \mathrm{M_\odot}}\) and \(\pqty{\kappa_R}_{1.4 \mathrm{M_\odot}}\), we study their distributions in Fig.~\ref{fig:5_distribution_histogram}.
In the case of \(\pqty{c_{sc}^2}_{1.4 \mathrm{M_\odot}}\), we know its value can be raised sharply when a quarkyonic phase is introduced in the core of the NS, and this leads us to hypothesise that the cause of the bimodality might be attributed to separate populations of NSs with and without a quarkyonic core.
As such, in Fig.~\ref{fig:5_distribution_histogram}, we plot the marginalized distributions from all three constrained sets by segregating the parameter vectors into hadronic (in green) and quarkyonic (in magenta) on the basis of the NS's central density being smaller or larger than \(\rho_t\).
We further note that in the correlation plots of both \(\pqty{c_{sc}^2}_{1.4 \mathrm{M_\odot}}\) and \(\pqty{\kappa_R}_{1.4 \mathrm{M_\odot}}\) with \(\pqty{\dv*{R}{M}}_{1.4 \mathrm{M_\odot}}\) and \(\Delta R\), the distributions look akin to being comprised of two lobes.
Fig.~\ref{fig:5_distribution_histogram} thus shows the marginalized posterior distributions of these four quantities.
It is evident from the figure that composition indeed causes two different distributions to appear for all the considered quantities, and the distributions are separable to a high degree for \(c_{sc}^2\) and \(\kappa_R\).
This indicates both that the presence of a quarkyonic phase in NSs can cause distinctive change in properties of the star, and that proper combinations of these properties can be potentially useful indicators for detecting the presence of a quarkyonic phase in NS\@.

\begin{figure}
	\centering
	\includegraphics[width=\linewidth, keepaspectratio]{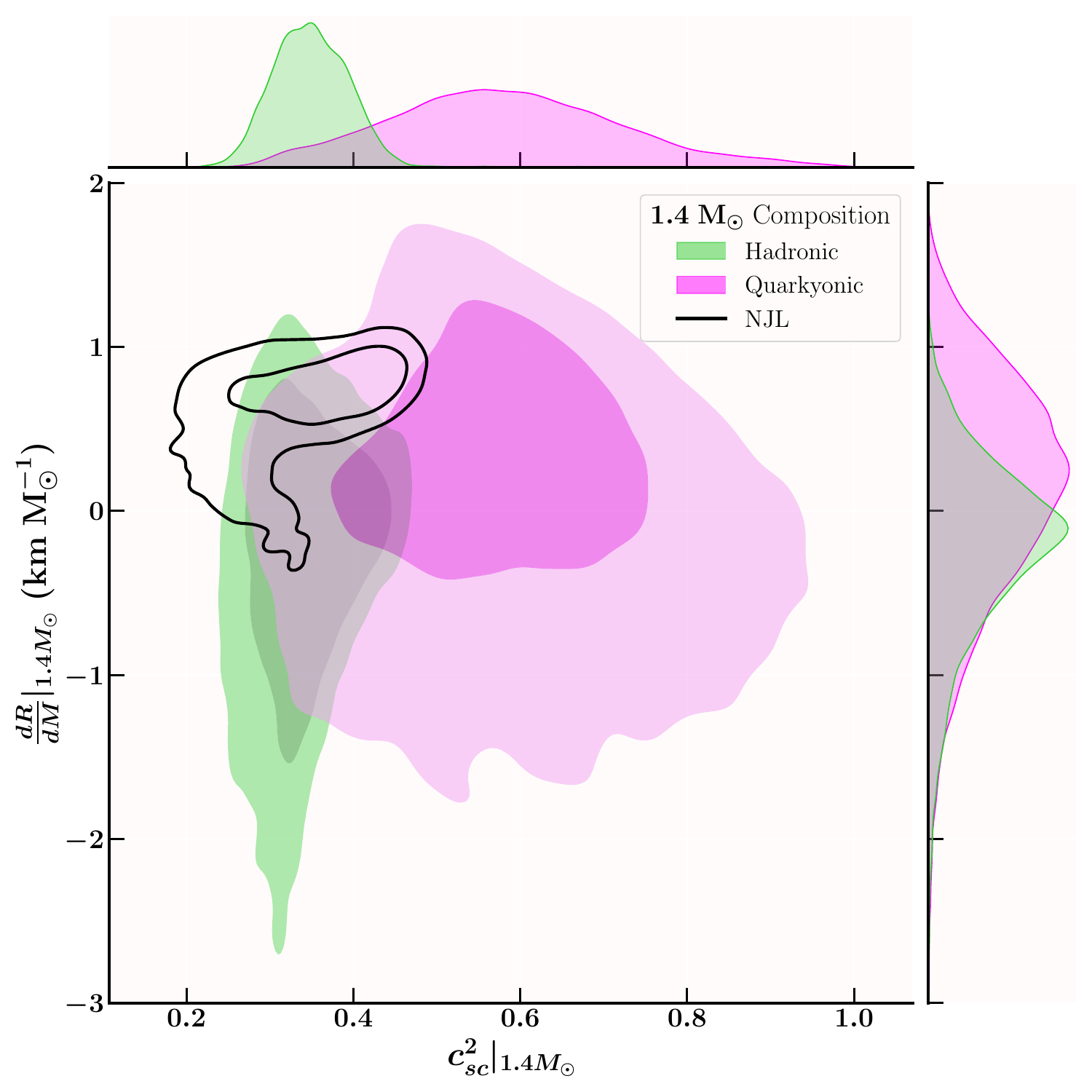}
	\caption{Slope of the \(R(M)\) relation plotted against the central sound speed for canonical mass NSs, with the distributions being coloured magenta (green) according to the presence (absence) of a quarkyonic phase in the canonical NS\@. The dark and light shadings denote the \(1\) and \(2\sigma\) regions, respectively. The histograms on the margin show the distributions of the two quantities. The contours outlined in black are the \(1\) and \(2\sigma\) regions for the NJL models of Ref.~\cite{Albino:2025puc}.}\label{fig:6_dRdM_cs2}
\end{figure}

We plot the marginalized posteriors of \(\pqty{\dv*{R}{M}}_{1.4 \mathrm{M_\odot}}\) and \(\pqty{c_{sc}^2}_{1.4 \mathrm{M_\odot}}\) against each other in Fig.~\ref{fig:6_dRdM_cs2} to test how well a combination of these two quantities would perform as an indicator for detecting the presence of a quarkyonic phase in canonical mass NSs.
The figure has the quarkyonic population in magenta and the hadronic in green, with the darker and lighter shaded regions outlining the \(1\) and \(2\sigma\) bounds.
Also, we plot the histograms from Fig.~\ref{fig:5_distribution_histogram} along the margins.
We can see from the figure that there does indeed exist separability to a certain degree between the two populations, but it arises mostly from the sound speed which rises sharply when transitioning to the quarkyonic phase.
Only considering the \(1\sigma\) regions we note that the quarkyonic population has a mostly positive value of the slope, but the hadronic population shows no such preference for the slope's value.
While this is an indicator of the populations being separable, it does not work well as a potential classifier for the detection of a quarkyonic core.
The figure also features the bounds obtained for the NJL models of Ref.~\cite{Albino:2025puc}, which are NSs with massive quark cores.
This region, shown in black, does not show agreement with either the hadronic or the quarkyonic population due to the model not being able to raise the speed of sound with sufficient rapidity.
It should be noted that the NJL model yielded \(R(M)\) slopes which are almost always positive at canonical mass.

\begin{figure}
	\centering
	\includegraphics[width=\linewidth, keepaspectratio]{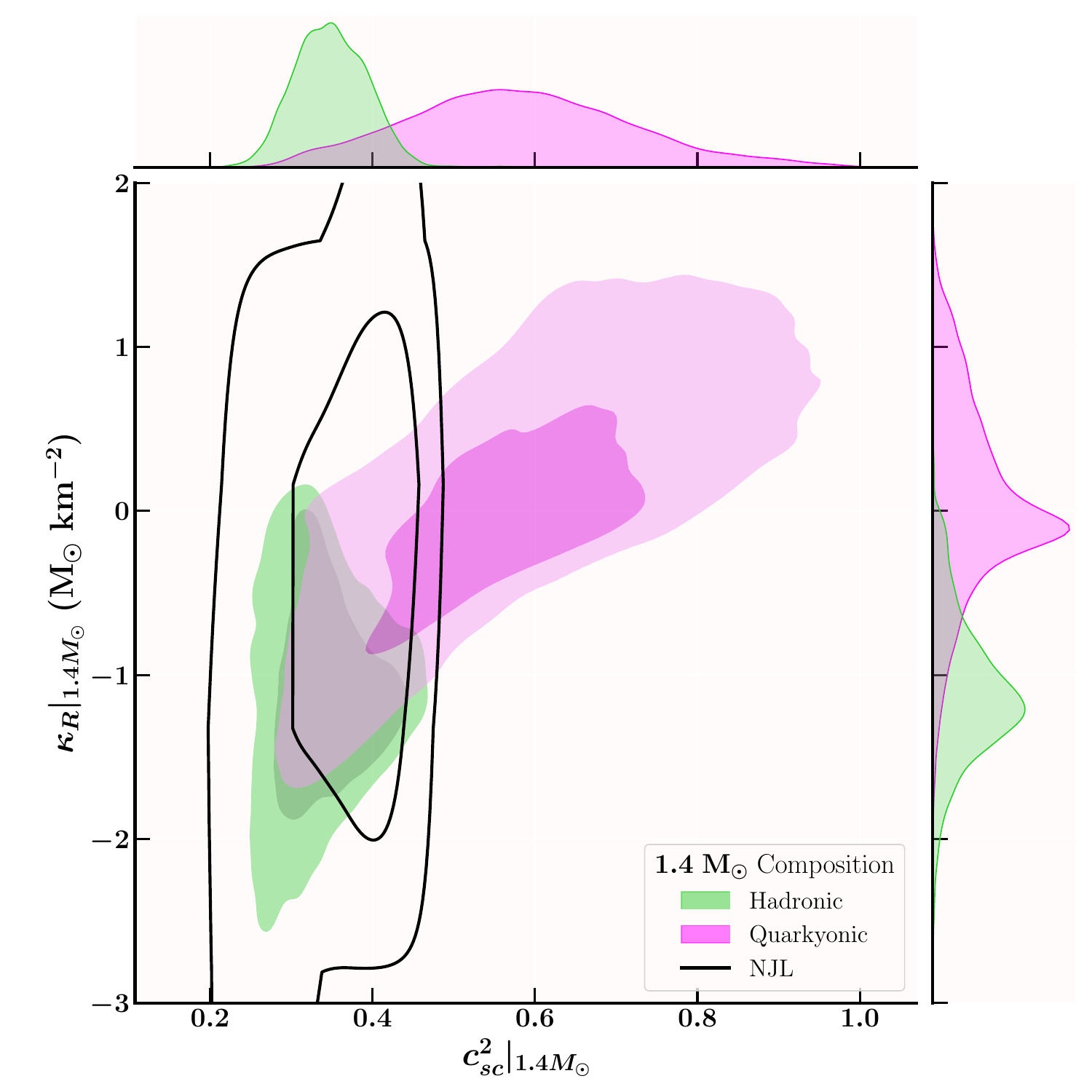}
	\caption{Curvature \(\kappa_R\) of the \(R(M)\) relation against the central sound speed for canonical mass NSs. The magenta (green) region indicates the presence (absence) of a quarkyonic phase in the canonical NS, with the light and dark regions denoting the \(2\) and \(1\sigma\) bounds, respectively. The histograms on the margin show the distributions of the two quantities. The contours outlined in black are the \(1\) and \(2\sigma\) regions for the NJL models of Ref.~\cite{Albino:2025puc}.}\label{fig:7_kappa_cs2}
\end{figure}

A better separation and a potentially better indicator can be obtained by combining \(\pqty{c_{sc}^2}_{1.4 \mathrm{M_\odot}}\) and \(\pqty{\kappa_R}_{1.4 \mathrm{M_\odot}}\), as we show in Fig.~\ref{fig:7_kappa_cs2}.
While this too shows some overlap, we note that here the \(1\sigma\) regions do not do so, and the separation between the quarkyonic and hadronic populations is distinctly visible.
Using these two quantities, which demonstrate the best distinction in their marginalized posteriors, provides a means to test the presence of quarkyonic matter.
This approach, along with the one in Fig.~\ref{fig:6_dRdM_cs2}, are however not practically feasible despite the promising nature of their results.
This is due to a current lack in our ability to determine the central sound speed of the star.
We thus need an observable quantity for use in identifying the presence of a quarkyonic phase.

\begin{figure}
	\centering
	\includegraphics[width=\linewidth, keepaspectratio]{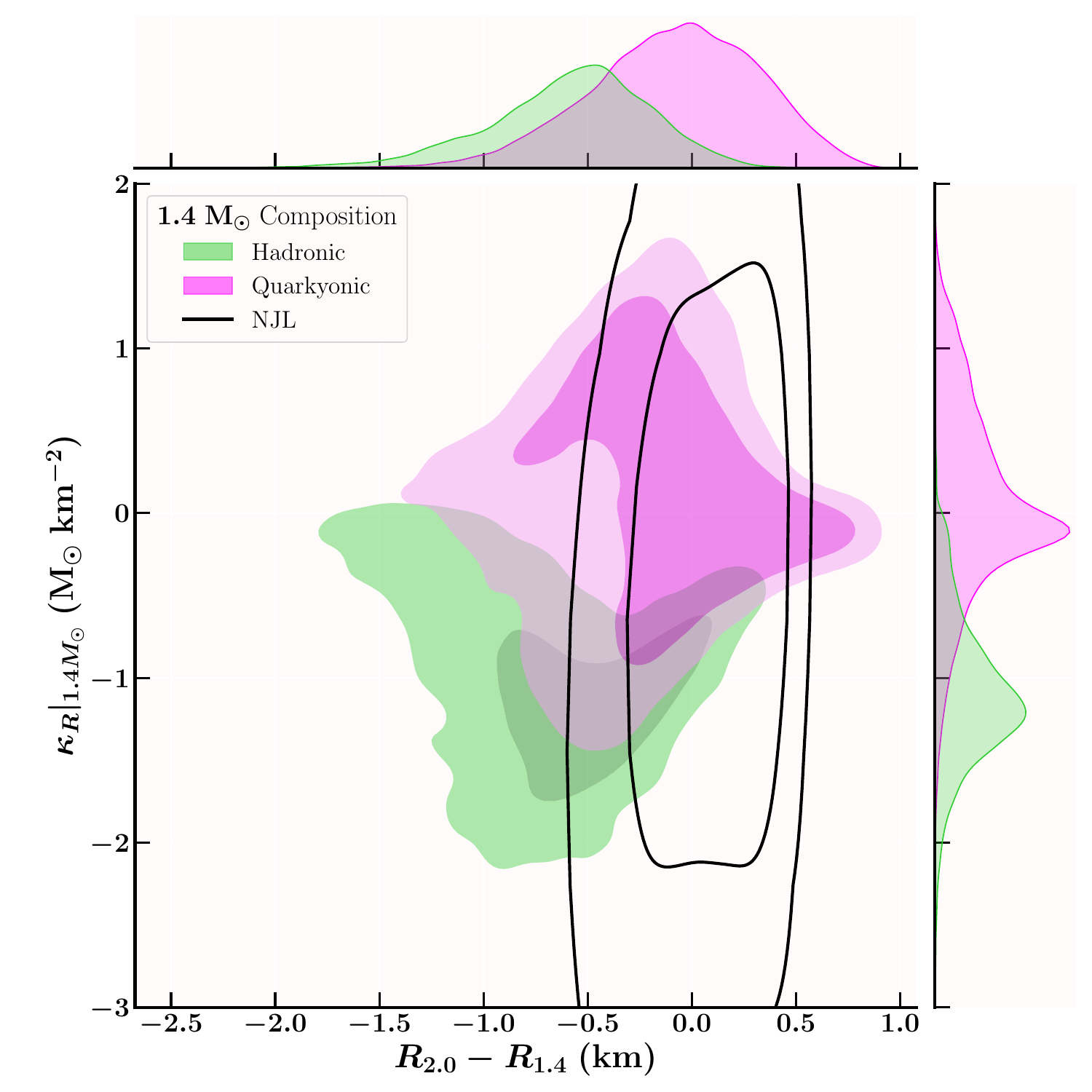}
	\caption{Curvature \(\kappa_R\) of the \(R(M)\) relation at canonical mass against the radial difference of \(2.0\ \mathrm{M_\odot}\) and \(1.4\ \mathrm{M_\odot}\) NSs. The magenta (green) region indicates the presence (absence) of a quarkyonic phase in the canonical NS, with the light and dark regions denoting the \(2\) and \(1\sigma\) bounds, respectively. The histograms on the margin show the distributions of the two quantities. The contours outlined in black are the \(1\) and \(2\sigma\) regions for the NJL models of Ref.~\cite{Albino:2025puc}.}\label{fig:8_kappa_deltaR}
\end{figure}

\(\Delta R = R_{2.0} - R_{1.4}\) being correlated to both the slope and the central sound speed, and showing a better separation in the marginalized distributions is thus the best candidate to use alongside the curvature for detection of quarkyonic matter.
This combination results in Fig.~\ref{fig:8_kappa_deltaR}, where we see that the populations have decent separation with minimal overlap between the \(1\sigma\) regions.
Crucially, these signatures can be tested in the near future, the next generation of gravitational wave detectors and radio telescopes promise significant improvements to both the mass and radii measurement precisions~\cite{sathyaprakash_cosmography_2010, sathyaprakash_scientific_2012, Begnoni:2025oyd, maggiore_science_2020, baibhav_lisa_2020, zhang_accuracy_2021, thorpe_lisa_2009, blaut_accuracy_2011} which can then be used to determine both the \(\pqty{\kappa_R}_{1.4 \mathrm{M_\odot}}\) and \(\Delta R\) values.
If a neutron star falls in the quarkyonic region of the plot, it would be evidence in favor of quarkyonic or quarkyonic-like phase being present.
This usage of the curvature and \(\Delta R\) is, to our knowledge, a novel and unexplored indicator that can provide insights into the neutron star core composition.

In both Fig.~\ref{fig:7_kappa_cs2} and Fig.~\ref{fig:8_kappa_deltaR}, we again show the results for the NJL models of Ref.~\cite{Albino:2025puc} using the black contours.
Both show the massive quark core NSs to occupy regions in the plots that contain quarkyonic as well as hadronic populations.
These plots, along with Fig.~\ref{fig:6_dRdM_cs2}, thus show that distinguishing between a quark and a quarkyonic cores might not be possible with the combination of properties we are considering, but it would still be an evidence for the existence of quarks in the NS cores.

\section{Discussions and Conclusion}
The Bayesian inference process shows that it is possible to generate physically valid \(\beta\)-equilibriated quarkyonic equations of state which satisfy all recent NICER, LIGO/Virgo and theoretical bounds, making it a valid description of dense matter which we might encounter at the interior of neutron stars.
From the validated equations of state, studying their macroscopic properties provides us with a rich trove of insights.
Importantly, we find that certain properties of NSs can be segregated on the basis of the star having a quarkyonic phase in the core, and these properties can be coupled together to serve as experimental probe into the composition of NS cores.
A large proportion of the observed NSs are in the vicinity of the canonical mass, making it more likely for the improved accuracy measurements from the future generation of detectors to also lie in this neighbourhood.
Also, as we move towards the maximum mass, the shape of the \(R(M)\) curves start becoming similar across compositions, making the properties we have identified not remain viable.
This leads us to restrict our study to the composition of canonical mass NSs.
The best combination of such properties, we find, is the central sound speed and the curvature of the \(R(M)\) curve at the canonical mass, and plotting them gives us a clean distinction between hadronic and quarkyonic cores.
The use of the central sound speed is, however, not a feasible method as we do not have a method to probe it.
While it might be possible to infer the value from asteroseismological means, or from EoS reconstruction from the \(R(M)\) relation, we would like to use properties that are more readily accessible.

The central sound speed being correlated to the radius difference of \(2\ \mathrm{M_\odot}\) and \(1.4\ \mathrm{M_\odot}\) NS (denoted by \(\Delta R\)), and it showing a reasonable separation between quarkyonic and hadronic populations makes it our property of choice.
A good separation is seen between the populations in the \(\kappa_R\)-vs-\(\Delta R\) plot, and an observation in the quarkyonic region of this plot would be a strong argument for the presence of quarkyonic phase in NS.
With the high accuracy measurements of mass and radii promised by the next generation of detectors, we would require only three (3) unique NS to be measured to obtain the curvature, thereby making it possible to infer the nature of NS core compositions from only their mass and radii.

Another observation we would like to note is that both we and~\citeauthor{Bauswein:2025dfg} have used \(\kappa_R\) as a probe into change in composition from the purely nucleonic case, with theirs being the appearance of hyperons in the system.
Hyperons appearing leads to softening of the EoS which caused \(\kappa_R\) to decrease, while we saw an increase in \(\kappa_R\) due to quarkyonic matter stiffening the EoS.
As observed by~\citeauthor{Bauswein:2025dfg}, we confirm that \(\kappa_R\) would be a better predictor of EoS stiffening than \(\dv*{R}{M}\).
Since the appearance of different particles beyond nucleons either stiffen or soften the EoS, \(\kappa_R\) can be a good indicator of such species appearing.
A large contrast is seen between the softening results of Ref.~\cite{Bauswein:2025dfg} and our stiffening results, suggesting that there is an even better chance of distinguishing these two behaviors than firmly establishing the validity of either one.
With this, we would like to mention the caveat that some of the results we have presented might possibly be replicated by other methods of stiffening the EoS, as were seen in the NJL model results.

\section{Acknowledgement}
PJK and BK acknowledge the usage of IUCAA HPC computing facility under project ID hpc2502001. T.M. gratefully acknowledges financial support from FCT – Fundação para a Ciência e Tecnologia, I.P., through national funds within the project UID/04564/2025 (DOI\@: 10.54499/UID/04564/2025). TZ acknowledges support by the Network for Neutrinos, Nuclear Astrophysics and Symmetries (N3AS) through the National Science Foundation Physics Frontier Center, Grant No. PHY-2020275. JML acknowledges funding from the US Department of Energy under Grant DE-FG02-87ER40317.

\bibliography{main}
\end{document}